\documentclass[twocolumn,preprintnumbers,amsmath,amssymb]{revtex4}

\usepackage{graphicx}

\begin{document}

\title{Weak Localization Thickness Measurements\\
        of Si:P Delta-Layers}

\author{D.F. Sullivan}
\email{dfs@lps.umd.edu}
\author{B.E. Kane}

\affiliation{Laboratory for Physical Sciences\\
University of Maryland, College Park, MD 20740}

\author{P.E. Thompson}
\affiliation{Naval Research Laboratory, Washington DC 20375}

\date{\today}

\begin{abstract}
We report on our results for the characterization of Si:P delta-layers grown by low temperature molecular beam epitaxy. Our data shows that the effective thickness of a delta-layer can be obtained through a weak localization analysis of electrical transport measurements performed in perpendicular and parallel magnetic fields. An estimate of the diffusivity of phosphorous in silicon is obtained by applying this method to several samples annealed at 850 $^{\circ}$C for intervals of zero to 15 minutes. With further refinements, this may prove to be the most precise method of measuring delta-layer widths developed to date, including that of Secondary Ion Mass Spectrometry analysis.
\end{abstract}

\maketitle

\section{}

The study of delta-layer ($\delta$-layer) structures is a very active area of research in contemporary semiconductor physics\cite{1}, primarily due to the device applications they offer, including ultra-high efficiency CCDs, low-energy particle detectors\cite{2} , and resonant interband tunnel diodes (RITDs)\cite{3} . In addition, recent proposals for the construction of a silicon-based quantum computer\cite{4} , in which accurately positioned phosphorous donors function as quantum bits (qubits), make the study of Si:P $\delta$-layers an obvious first step in the direction of achieving this goal\cite{5} . This letter will demonstrate that an analysis of weak localization signals in perpendicular and parallel magnetic fields provides a very precise method for measuring the thickness of $\delta$-layers. This quantity is extremely important for characterizing the quality of these structures, since broadening can occur during growth (segregation) as well as in subsequent processing (diffusion). At present perhaps the most common method for determining the thickness of $\delta$-layers is through the use of Secondary Ion Mass Spectrometry (SIMS). Unfortunately, SIMS has several drawbacks which limit its usefulness in this context. One is that SIMS is a destructive technique. More importantly, the spatial resolution of SIMS is limited\cite{6}  to approximately 5 nm. A more sensitive probe is necessary for applications requiring ultra-thin (near monolayer) distributions. Our technique is free of these limitations. 

When the two-dimensional electron gas (2DEG)\cite{7} associated with a $\delta$-doping distribution is cooled to low temperatures (T$\leq$ 4.2K), the resistance of a device fabricated from such a structure decreases in an applied magnetic field ($\mathbf{B}$). This behavior is characteristic of the phenomenon of weak localization \cite{8} (WL), a quantum effect due to the Aharanov-Bohm interference between oppositely directed and closed electron paths in the device. At $\mathbf{B}$=0, the amplitudes associated with these paths are coherent, interfere constructively, and make the device more resistive. For $\mathbf{B}$$\neq$0, the interference is destructive, and the sample is less resistive. An applied parallel $\mathbf{B}$ is consequently an extremely sensitive probe of the effective thickness of the electron system under study. An early WL \cite{9} study on Si MOSFETs (Metal Oxide Semiconductor Field Effect Transistors) in parallel fields measured the root-mean-square (rms) thickness of the 2DEG with sub-monolayer resolution ($\sim$ 0.2 nm). If such resolution could be obtained in $\delta$-layer studies, it would better SIMS capabilities by roughly an order of magnitude. With the caveat that the sample must be electrically conducting at low temperatures for this technique to work (meaning for Si:P a doping density greater than the metal-insulator transition (MIT) value of $\sim$ $4x10^{18} (cm^{-3})$), we see no fundamental reason why monolayer resolution should not be achievable.

 A SIMS profile from the wafer used in our experiments is shown in Fig. 1. This sample was grown by low temperature molecular beam epitaxy in a six step process. The P $\delta$-layer itself was grown at 320 $^{\circ}$C, "sandwiched" between Si grown at higher temperatures. This $\delta$-distribution has a full width at half maximum (FWHM) of approximately 5 nm, close to the resolution limit of SIMS. For electrical measurement, four 50 $\mu$m x 50 $\mu$m van der Pauw devices were fabricated using standard lithographic techniques, three of which were subjected to an 850 $^{\circ}$C anneal for 5, 10 and 15 minutes, respectively, to broaden the $\delta$-layer by thermal diffusion (henceforth these devices will be referred to by annealing time as cook 0, cook 5, cook 10 and cook 15). These samples were then mounted in a dilution refrigerator fitted with a tilting stage which enabled adjustment of the angle ($\theta$) between the sample plane and $\mathbf{B}$. The base temperature of this refrigerator was 130 mK, and from temperature-dependent resistivity measurements we estimate that the relevant electronic temperature was $\sim$200 mK. Resistivity measurements were performed using a resistance bridge, four-wire measurement techniques and the appropriate Van der Pauw transformation. The signals induced in the samples due to a current bias ($\sim$ 1µA) were preamplified and recovered using low-frequency lock-in detection. We determined the orientation of our samples through careful measurements of the Hall coefficient, $\rho_{xy}$ = Bsin($\theta$)/ne (n is the sample carrier density, e the electron charge). By taking successive Hall traces very close to parallel field we were able to determine $\theta$ to better than 0.05$^{\circ}$ (this stray $\mathbf{B}$ would contribute only $\sim$0.2 nm to our thickness estimates determined below). In addition, our relatively small sample size minimized the influence any $\mathbf{B}$ inhomogeneities may have had on our measurements.

\begin{figure}
\includegraphics[width=3.125in]{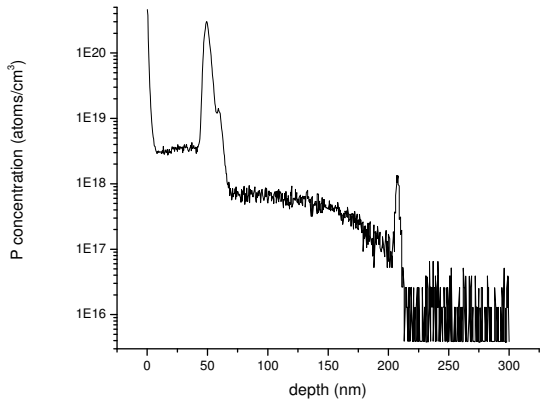}
\caption{A SIMS profile of an Si:P $\delta$-layer sample. The $\delta$-layer is located approximately 50 nm below the surface, and has a FWHM of $\sim$ 5 nm.}
\end{figure}

For a two-dimensional system the WL signal depends on both the magnitude and direction of the applied $\mathbf{B}$. When $\mathbf{B}$ is perpendicular to the $\delta$-layer ($\mathbf{B_{\perp}}$), the change in conductance for a system with weak spin and spin-orbit scattering (as is the case for Si:P) is given by\cite{10} :

\begin{eqnarray}
   \delta\sigma(B_{\perp})= (\frac{e^2}{2{\pi}^2\hbar})[\Psi(\frac{1}{2}+ \frac{\hbar}{4eB_{\perp}{L_\phi}^2})-\\
\Psi(\frac{1}{2}+ \frac{\hbar}{2eB_{\perp}{L}^2}) + \ln(\frac{2{L_\phi}^2}{{L}^2})\nonumber]
\end{eqnarray}	     

Here $\Psi$ is the digamma function, \(\L_{\phi}\) the dephasing length (the mean distance over which the wave function loses phase coherence), and L the mean free path. In parallel fields ($\mathbf{B_{\parallel}}$)
 the change in conductance becomes a logarithmic correction, $\emph{quadratic}$ in $\mathbf{B_{\parallel}}$\cite{11} and characterized by a single parameter $\gamma$:      

\begin{equation}
   \delta\sigma(B_{\parallel})= (\frac{e^2}{2{\pi}^2\hbar}\\)\ln(1 + \gamma{B_{\parallel}^{2}}).
\end{equation}

The data and theoretical fits for cook 0 are shown in Fig. 2, along with the parallel field plots for the annealed samples. The difference between the $\mathbf{B_{\perp}}$ and $\mathbf{B_{\parallel}}$ configurations can be understood by the significantly reduced flux penetrating the $\delta$-layer when $\mathbf{B}$ and the current are coplanar.

\begin{figure}
\includegraphics[width=3.125in]{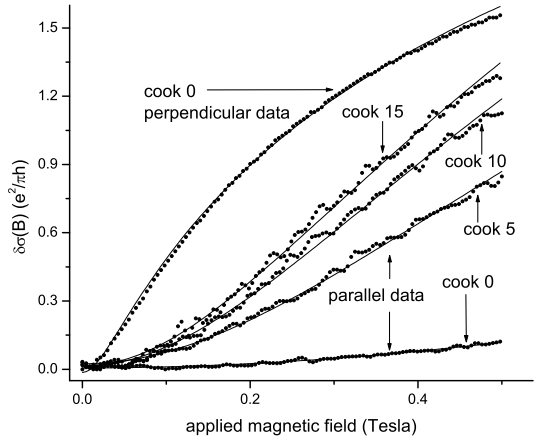}
\caption{Weak localization data (points) with fits (solid lines) to equations (1) and (2) in perpendicular and parallel magnetic fields. The conductance changes much more rapidly with applied field in the perpendicular configuration due to a larger magnetic flux through the device.}
\end{figure}

\scriptsize
\begin{table}
\scriptsize
\caption{\label{tab:table1}Experimental results for $\delta$-layer samples annealed ("cooked") at 850 $^{\circ}$C for intervals of 0 to 15 minutes. The parameters L and $L_{\phi}$ were extracted from fits to Eqn. (1), $\gamma$ from fits to Eqn. (2). The mean-square thickness of a given layer is proportional to the parallel field parameter $\gamma$. The dominant error in the thickness estimate comes from the mean free path, L.  }
\begin{tabular}{|c|c|c|c|c|c|c|c|}\hline
sample & resistivity     & electron		 & mean 	& dephasing  	& $\gamma$    & mean \\
       &	         & density 		 & free 	& length     	& 	      &thickness \\
       &                 &   (n)   		 & path 	& ($L_{\phi}$)	&   	   	& (T)      \\
       &		 &    	   		 &  (L) 	&		&      	 	 &          \\
       &[$\Omega$/$\Box$]& [$10^{14}/cm^{2}$]	&[nm]		&[nm]		&[$1/T^{2}$]	&[nm]\\\hline
	 & 	   	 &       		&    30		& 101		 & 	 	&14  \\ 
cook 0	&	428.6		&	1.40         	&	+/-	 & +/-       	 &   0.49     	&  +/-     \\
	&		 &           		&	13   	 & 3     	 &           	&     3    \\\hline
	 & 	   	 &       		&    33		& 117		 & 	 	&42\\ 
cook 5	&	300.9	&	1.39        	&	+/-	 & +/-       	 &   5.31	&  +/-     \\
	&		 &           		&	14  	 & 11   	 &           	&     10  \\\hline

	 & 	   	 &       		&    37		&124		 & 	 	&	51\\ 
cook 10	&	267.9	&		1.42   	&	+/-	 & +/-       	 &   7.54 	&  +/-     \\
	&		 &           		&	13   	 & 3     	 &           	&     9   \\\hline

	 & 	   	 &       		&    38		&127	 & 	 	&	61\\ 
cook 15	&	251.6	&		1.49  	&	+/-	 & +/-       	 &   11.25&	+/-     \\
	&		 &           		&	14  	 & 9    	 &           	&     12 \\\hline
\end{tabular}
\end{table}
\normalsize

To extract an rms 2DEG thickness from the above fitting parameters, we employ a theoretical model developed for MOSFETs \cite{12} in the lowest subband (as far as we are aware, a model of weak localization in $\delta$-layers does not exist at this time). Of particular concern is how the spreading of the donor distribution affects the magnetoresistance. As described in Ref. 12, a surface can be characterized by an rms amplitude (here corresponding to, T, the $\delta$-layer thickness), and a length $\L_{c}$ over which the fluctuations in the surface height are correlated. Different thickness estimates are obtained depending on the relative magnitudes of L, $\L_{\phi}$, and $L_{c}$, with analogs to homogeneous ($\L_{c}$$\ll$$\L_{\phi}$) and inhomogeneous ($\L_{c}$$\gg$$\L_{\phi}$) broadening in nuclear magnetic resonance. For $\delta$-layers, the roughness likely arises from fluctuations in the positions of individual donors, suggesting a short correlation length of order the mean donor spacing $\L_{c}$ = 1/$\sqrt{n}$ ($\sim$1 nm).  We have no way to independently measure $\L_{c}$, so we make this approxiamtion and thus have the inequality (see Table I) $\L_{c}$$\ll$L$\ll$$\L_{\phi}$. This leads to an estimate of the $\delta$-layer thickness given by (see Eqn. 55 of Ref. 12):

\begin{equation}
   T = {(\frac{1}{4\pi}\\)}^{1/4}{[{(\frac{\hbar}{eL_{\phi}})}^{2}(\frac{L}{L_{c}})\gamma]}^{1/2}.
\end{equation}

 Using Eqns. (1)-(3) we have calculated the rms thickness of the $\delta$-layer in each of our devices, and shown our results in Table I. Our expectation was that the mean square thickness of a sample would be directly proportional to the time it was annealed. Table I clearly establishes this relationship, thereby verifying that this technique does indeed measure $\delta$-layer thicknesses. In Fig. 2 we observe that the parallel field parameter $\gamma$ (and therefore the rms thickness) systematically increases with annealing time. Using these results we can estimate the diffusivity, D, of Si:P at 850 $^{\circ}$C. Assuming an initially Gaussian profile with thickness $T_{0}$, the mean-square thickness after annealing for a time $\Delta$t is expected \cite{13} to be \newline $T^{2} = T_{0}^{2} + 2D\Delta t$. Our data is plotted in Fig. 3. We have rejected the cook 0 data point from our linear fit, since it is well known that defects such as interstitials and vacancies enhance diffusion\cite{14} . Considering the low temperature under which our $\delta$-layer was grown (320 $^{\circ}$C), it seems likely that a large number of defects were present prior to annealing, and that this disorder was subsequently "cooked out" (many examples of this sort have been discussed in the literature \cite{15}). Fitting a line to the annealed data points, we obtain a slope of 186($nm^{2}$/min). From this we calculate the diffusion coefficient of Si:P at 850 $^{\circ}$C to be $D = 1.6x10^{-14}(cm^{2}/sec)$. Another study \cite{16} of Si:P using SIMS measurements reports that for this temperature $D=4.16x10^{-16}(cm^{2}/sec.)$. This discrepancy may stem from a number of sources, including the thickness model we have employed being inadequate (particularly 1/$\sqrt{n}$ underestimating $\L_{c}$), different sample preparation techniques or perhaps systematic effects associated with $\mathbf{B}$ (such as imperfect sample alignment or field inhomogeneities). 

\begin{figure}
\includegraphics[width=3.125in]{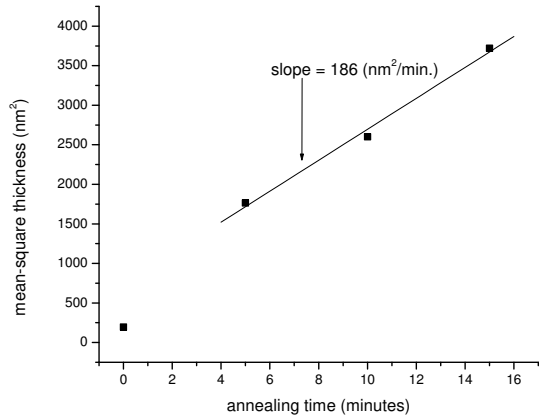}
\caption{A plot of the mean square $\delta$-layer thickness versus annealing time at 
850 $^{\circ}$C. The slope of the line fit to the annealed data points yields an estimate of the diffusion coefficient of Si:P at this temperature: 
$D = 1.6x10^{-14}(cm^{2}/sec)$.}
\end{figure}

Based on our diffusivity results and SIMS data, it seems clear that we have systematically overestimated the thicknesses of the $\delta$-layers in our samples. We have, however, shown a monotonic increase in a devices rms $\delta$-layer thickness subsequent to annealing. Therefore we believe that by doing more extensive measurements and eliminating the various systematic errors in our experiments (whose contribution to our thickness estimates we have not included), this technique will provide the most precise method of measuring the thickness of very thin $\delta$-layers yet developed. For the future, we plan to investigate samples with densities near the MIT limit, which should give larger WL signals and perhaps be more tractable from a theoretical viewpoint.

   We would like to thank Richard Webb and Carlos Sanchez for helpful advice and discussions, and the National Security Agency and ARDA for financial support. The work at NRL was supported by the Office of Naval Research.

\end{document}